\documentclass[11pt,fleqn]{article}
\usepackage{graphicx}
\usepackage{amsmath,amssymb}
\usepackage{amsfonts,color}

\usepackage{amsfonts,amssymb,cite}
\usepackage{graphicx}

\def\noi{\noindent}

\newcommand{\Title}[1]{\noi {{\Large\bf #1}}\\[1ex]}

\def\Aunames#1{\noi{\bf #1}}
\def\au#1{${}^{#1}$}
\def\Addresses#1{\medskip\noi \protect
	\begin{description}\itemsep -3pt {\it #1} \end{description}}
\def\adr#1#2{\item[${}^{#1}$]{\it #2}}

\newcommand{\Abstract}[1]{\vskip 2mm \begin{center}
        \parbox{16.4cm}{\small\noi #1} \end{center}\medskip}

\def\email#1#2{\footnotetext[#1]{e-mail: #2}\addtocounter{footnote}{1}}

\def\nqq{\hspace*{-2em}}

\usepackage{color}
\def\red#1{{\color{red} #1}}
\def\blue#1{{\color{blue} #1}}

\def\Jl#1#2{#1 {\bf #2},\ }

\def\ApJ#1 {\Jl{Astroph. J.}{#1}}
\def\CQG#1 {\Jl{Class. Quantum Grav.}{#1}}
\def\DAN#1 {\Jl{Dokl. AN SSSR}{#1}}
\def\GC#1 {\Jl{Grav. Cosmol.}{#1}}
\def\GRG#1 {\Jl{Gen. Rel. Grav.}{#1}}
\def\IJMPD#1 {\Jl{Int. J. Mod. Phys. D}{#1}}
\def\JETF#1 {\Jl{Zh. Eksp. Teor. Fiz.}{#1}}
\def\JETP#1 {\Jl{Sov. Phys. JETP}{#1}}
\def\JHEP#1 {\Jl{JHEP}{#1}}
\def\JMP#1 {\Jl{J. Math. Phys.}{#1}}
\def\NPB#1 {\Jl{Nucl. Phys. B}{#1}}
\def\NP#1 {\Jl{Nucl. Phys.}{#1}}
\def\PLA#1 {\Jl{Phys. Lett. A}{#1}}
\def\PLB#1 {\Jl{Phys. Lett. B}{#1}}
\def\PRD#1 {\Jl{Phys. Rev. D}{#1}}
\def\PRL#1 {\Jl{Phys. Rev. Lett.}{#1}}

\def\lal{&&\nqq {}}

\def\beq{\begin{equation}}
\def\eeq{\end{equation}}
\def\bear{\begin{eqnarray}}
\def\bearr{\begin{eqnarray} \lal}
\def\ear{\end{eqnarray}}
\def\earn{\nonumber \end{eqnarray}}

\def\ep{\epsilon}

\begin{document}
\twocolumn[
\thispagestyle{empty}
\bigskip\bigskip

\Title{A note on ``Traversable wormholes in Einstein-Dirac-Maxwell theory''}

\Aunames{S. V. Bolokhov,\au{a,1} K. A. Bronnikov,\au{a,b,c,2} 
			Serguey Krasnikov,\au{d}	 and M. V. Skvortsova\au{a,3}}

\Addresses{\small
\adr a {Peoples' Friendship University of Russia (RUDN University),\\ 
		ul. Miklukho-Maklaya 6, Moscow 117198, Russia}
\adr b {Center for Gravitation and Fundamental Metrology, VNIIMS, 
		Ozyornaya ul. 46, Moscow 119361, Russia}		
\adr c {National Research Nuclear University ``MEPhI'', 
		Kashirskoe sh. 31, Moscow 115409, Russia}
\adr d	{Central Astronomical Observatory at Pulkovo, St. Petersburg, 196140, Russia}
        }	


\Abstract
  {In their Letter  [Phys. Rev. Lett. {\bf 126}, 101102 (2021); arXiv: 2010.07317], 
  J. L. Bl\'azquez-Salcedo, C. Knoll and E. Radu have constructed a very interesting 
  class of wormhole solutions in general relativity (GR), supported by a pair of classical 
  charged spinor fields obeying the Dirac equation. The main new feature of these solutions 
  is that such Dirac spinor fields can possess exotic properties, necessary for the existence 
  of static wormhole configurations in GR.
  The present note contains a few remarks clarifying some points concerning this approach.}
  
\bigskip

]  
\email 1 {boloh@rambler.ru}
\email 2 {kb20@yandex.ru} 
\email 3 {milenas577@mail.ru}

{ 
\def\ep{{\epsilon}} 
\def\bh{black hole}
\def\bhs{black holes}
\def\wh{wormhole}
\def\whs{wormholes}
\def\asflat{asymptotically flat} 
\def\red#1{{\color{red} #1}}
\def\blue#1{{\color{blue} #1}}

{\bf 1.} The authors of \cite{pap} insist that their spinor fields are particle wave functions,
  which entails their one-particle normalization and the corresponding values of the solution parameters. 
  
  As shown by Fig.\,2 in \cite{pap}, the curvature scalars at the throat are of the order of unity 
  in Planck units (the Ricci and Kretschmann scalars are $\approx 0.1$ and $\approx 6$, respectively).
  This suggests --- as no fine tuning is mentioned in the text --- that vacuum polarization inducing
  geometric corrections to the Einstein equations are of the same order of magnitude as the 
  their left-hand sides themselves. Thus the metric solving the Einstein equations generally does 
  not solve the actual equations of motion.

  On the other hand, a direct $n$-particle generalization of the present solutions is impossible due to 
  the Pauli exclusion principle, so the problem should be attacked anew.

{\bf 2.} The authors claim that if $\nu' \ne 0$ at the throat (where $2\nu = \ln(-g_{tt})$) requires
  the presence of extra matter forming a thin shell at the throat $r=0$. It is not true: in fact, 
  $\nu'(0) \ne 0$ simply means that the \wh\ must be nonsymmetric relative to the throat, while 
  a thin shell is only required if one postulates an unnecessary $Z_2$ symmetry with respect 
  to $r=0$. A search for nonsymmetric \wh\ solutions can 
  be one of possible interesting extensions of the authors' study. 
  Examples of other non-$Z_2$-symmetric spherical \whs\ are well known. 
  
{\bf 3.}  It is asserted in \cite{pap} that the solutions obtained numerically are $Z_2$-symmetric.
   Meanwhile,  
  after Eq.\,(8) we read that ``$Q_N$ is the Noether charge of a spinor (or number of particles),'' 
  and then, after the unnumbered expression for $Q_N$, we read that  ``Similar relations
  hold for the $r < 0$ region, with mass, electric charge and Noether charge changing sign.''
  This raises some objections: (i) it looks too strange to have a negative particle number, 
  not to mention that a normalizing integral should cover the whole volume, $r \in {\mathbb R}$, 
  and the particle number should pertain to the whole configuration; (ii) $Z_2$ symmetry 
  implies equal ADM masses on both ends (as is confirmed by the curve for $g_{tt}$ in Fig.\,2), 
  then how to understand a ``mass changing sign''?
  
{\bf 4.} The metric (5), obtained as an exact solution with massless spinors, has been previously 
  discussed as a possible metric in a brane world supported by a bulk-induced tidal stress-energy 
  tensor \cite{kb1, kb2}.

\newpage  

{\bf 5.} The words in the abstract, that the \wh\ solutions are obtained ``without needing any form of 
  exotic matter,'' look misleading since exotic matter (by definition, matter violating the Null 
  Energy Condition) is quite necessary at a \wh\ throat  \cite{morris}. It would be better to say that 
  Dirac spinor fields become exotic matter under certain circumstances.
  
  We conclude that the authors of \cite{pap} have made an interesting contribution to \wh\ physics, 
  but hope that our remarks clarify some important points concerning their approach.

  It is worth mentioning that the idea of using spinor fields to construct traversable wormholes
  in four dimensions was studied in a number of earlier works, for instance, by J. Maldacena
  et al.~\cite{malda}. 
  
  Since the present note was written and submitted (April 2021), a few relevant papers
  have appeared~\cite{radu2, roman1, roman2, wald}. In particular, J. L. Bl\'azquez-Salcedo
  et al.~\cite{radu2} presented a more detailed description of the wormhole solution under 
  discussion, including an investigation of the domain of existence for the solutions. 
  R. Konoplya et al.~\cite{roman1, roman2} obtained and discussed other wormhole 
  configurations supported by a pair of spinor fields, mentioning that changing the sign
  of the fermionic charge density at the throat can hardly be understood as a physically realistic
  scenario. Lastly, in the recent paper by D. Danielson et al.~\cite{wald} it is argued that
  the wormhole solution presented in~\cite{pap} suffers from a failure of the Maxwell and Dirac 
  fields to satisfy the necessary matching conditions.

} 

\end{document}